\documentclass[aip,amsmath,jmp,amssymb,preprint]{revtex4}
\usepackage{appendix}
\usepackage{graphicx}
\usepackage{natbib}

\begin{document}
\title{Cryogenic Microwave Imaging of Metal-Insulator Transition in Doped Silicon}
\author{Worasom Kundhikanjana}
\affiliation{Department of Applied Physics and Geballe Laboratory
for Advanced Materials, Stanford University, Stanford, CA 94305}
\author{Keji Lai}
\affiliation{Department of Applied Physics and Geballe Laboratory
for Advanced Materials, Stanford University, Stanford, CA 94305}
\author{Michael A. Kelly}
\affiliation{Department of Applied Physics and Geballe Laboratory
for Advanced Materials, Stanford University, Stanford, CA 94305}
\author{Zhi-Xun Shen}
\affiliation{Department of Applied Physics and Geballe Laboratory
for Advanced Materials, Stanford University, Stanford, CA 94305}

\date{\today}

\begin{abstract}
We report the instrumentation and experimental results of a cryogenic scanning microwave impedance microscope. The microwave probe and the scanning stage are located inside the variable temperature insert of a helium cryostat. Microwave signals in the distance modulation mode are used for monitoring the tip-sample distance and adjusting the phase of the two output channels. The ability to spatially resolve the metal-insulator transition in a doped silicon sample is demonstrated. The data agree with a semi-quantitative finite-element simulation. Effects of the thermal energy and electric fields on local charge carriers can be seen in the images taken at different temperatures and DC biases.
\end{abstract}
\maketitle 

\section{Introduction}

Variable-temperature (T) scanning probe microscopes (SPM) have become major scientific tools for modern condensed matter physics research. The cryogenic scanning tunneling microscope, for example, has tremendously advanced our understanding of the nature of high-temperature superconductors\cite{c1}. To obtain information other than the tunneling density of states, many low-T SPMs, such as the electrostatic force microscope \cite{c2,c3} and the scanning gate microscope \cite{c4}, have also been developed. These tools study the local electronic properties at low frequencies (less than 1 MHz), where Ohmic contacts to the sample are usually required. For high frequencies, a variable-T near-field scanning optical microscope has been implemented to perform dielectric imaging on transition metal oxides without the need of DC electrodes \cite{c5}. The high energy of optical photons, however, may cause undesired inter-band transitions that obscure the underlying physics. Electrical imaging in the microwave regime \cite{c6,c7} down to helium temperature is therefore important for fundamental research on complex materials and phase transitions under various conditions.

In this paper, we detail the operation of a cryogenic microwave impedance microscope (MIM) and the experimental results on a doped silicon (Si) sample. The application of this variable-T MIM on a strongly correlated material system has been reported elsewhere\cite{c8} and we will focus here on the instrumentation and a classical problem in semiconductor physics -- localization and activation of dopant carriers. The MIM setup is located in the exchange gas tube of a flow cryostat. Cantilever based probes with  shielded metal center conductors are employed. The scanning stages are capable of moving the sample in three dimensions in both coarse and fine steps. During the tip-sample approach, a small modulation voltage is applied to the z-piezo\cite{c9,c10} for locating the sample surface\cite{c11, c12, c13}. The ability of impedance imaging is demonstrated by the results on a patterned Si sample\cite{c13, c14, c15, c16} with impurity levels above the degenerate point. The expected MIM response calculated from the doping profile and the finite element modeling\cite{c17} (FEA) agree with the experimental results. The metal-insulator transition in doped Si is vividly visualized in the spatially resolved conductivity maps at different temperatures and DC biases.

\section{Experimental Setup}

\begin{figure}
	\centering
		\includegraphics[width=3.2in]{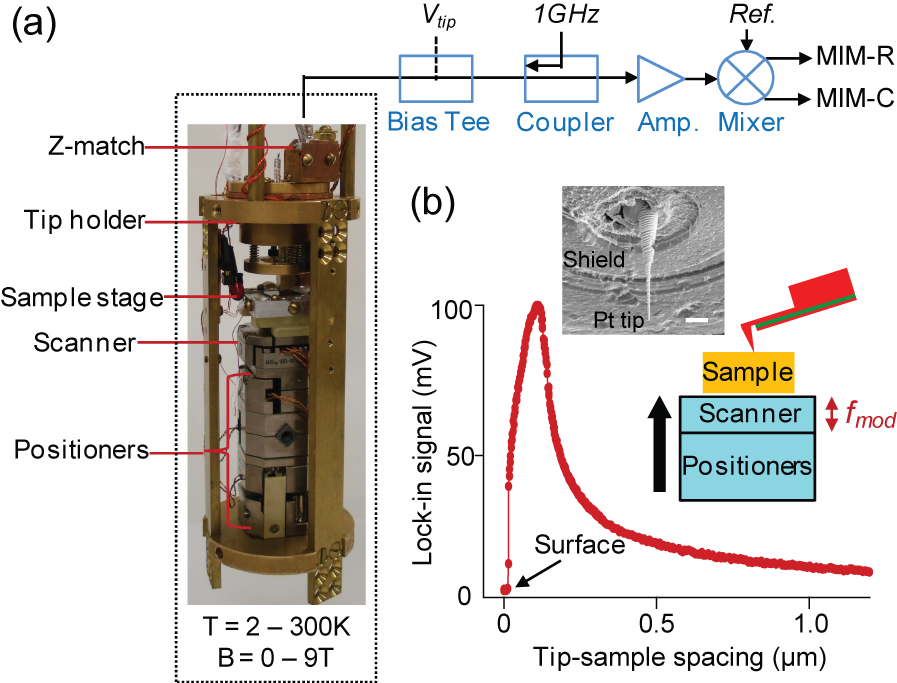}
		\caption{(color online) (a) Schematic of the low temperature MIM setup, including a picture of the tip-sample assembly and the block diagram of the microwave electronics. The distance modulation is turned on for tip-sample approaching and off for scanning. (b) A typical approaching curve, in which the lock-in reading is recorded as a function of the tip-sample spacing (z-scanner voltage). The inset shows schematically the tip-sample distance modulation. The shielded cantilever probe is employed and the SEM picture of the pt tip is shown. The scale bar equals to 2 $\mu$m. }
			\label{fig1}
\end{figure}

Fig. \ref{fig1}a shows the schematic of the cryogenic MIM setup. The reflected microwave signal from the tip is sent to the room temperature microwave electronics\citep{c18} through a $\lambda$/4 impedance-match section\citep{c14}. The conductive path on the cantilever is shielded up to the metal tip, which is made by focused-ion beam (FIB) deposition of Pt with a typical  apex of 100 - 200 nm (Fig. \ref{fig1}b inset). The layer structure of the microwave probe is described in Ref [11]. The current microwave electronics are optimized for operating at 1 GHz, while in principle this setup can operate at frequencies between 0.1 - 10 GHz. The sample stage underneath the probe is mounted on an Attocube assembly (Attocube Systems AG), which moves the sample with respect to the tip. Three positioners, two (ANPx100) for the xy directions and one (ANPz100) for the z direction, are used for coarse motion. Each positioner is equipped with a resistive encoder for position readout. The available travel range inside the cryostat is 3mm $\times$ 3mm in the xy-plane and 6 mm in the z-direction.The xyz-scanner (Attocube ANSxyz100) is used for the fine xy-scan and the tip-sample approach. The maximum scan range is 30$\mu$m $\times$ 30$\mu$m at 4 K.  This assembly is attached at the end of the exchange gas insert of a flow cryostat (Janis Research Company, Inc.) equipped with a 9 T superconducting magnet, allowing the sample temperature to be varied from 2 to 300 K. The cryostat is installed on an air table for vibration isolation.

The tip-sample approach is performed by a procedure similar to that described in Ref[10,11,13], in which the microwave signal itself is used for monitoring the tip-sample distance ($d$). A low-f (e.g. 400 Hz) AC voltage is applied to the z-scanner, resulting in a small sinusoidal modulation (amplitude $\sim$ 10 nm) of the tip-sample spacing. The MIM signals, also modulated at this frequency, are detected by lock-in amplifiers (Fig.  \ref{fig1}a). Since the signal line on the cantilever is well shielded, the demodulated microwave signals are essentially zero at large $d$. The signals become detectable when $d$ $ < $ 10 $\mu$m, rise sharply at $d$ $\sim$ 1 $\mu$m and drop immediately when the tapping motion is stopped by tip-sample contact (Fig. \ref{fig1}b). In the actual experiments, the sample is brought toward the tip in a stepper fashion by alternating motions of the z-scanner and the z-positioner. First, the z-positioner raises the sample stage until discernible lock-in signals are measured.  Then the z-scanner extends to look for the tip-sample contact point. If the z-scanner reaches the end of its travel before making the contact, it is fully retracted and the z-positioner moves half of the z-scanner range. The z-scanner then takes over again. The cycle is repeated until the tip-sample contact is established.  This procedure prevents possible tip damage from the abrupt z-positioner motion. The tip-sample approach is usually performed at multiple points along the edges of the scanning area. The z-scanner values at these points are recorded for a first order plane-fit of the sample surface. Finally, the z-modulation is turned off before the open-loop xy-scan. The compliance of the cantilever probe and the compensation of the sample tilt both help to minimize the tip-sample contact force.  Thus, even without a feedback mechanism, a tip can be used for extensive scans on relatively flat samples.  Due to technical complications, we have not attempted to implement a close-loop feedback mechanism, such as optical beam bounce \cite{c19}, tuning fork \cite{c20} or piezo-resistive self-sensing \cite{c4}, in this system. Implementation of a topography servo will be the subject of our future work as we improve the probe design. 

The distance modulation process also plays an important role for setting the phase of the microwave electronics.  MIM measures the effective tip-sample impedance during the scan. The phase of the reference signal needs to be defined such that the two orthogonal outputs, MIM-R and MIM-C, correspond to the real and imaginary parts of this impedance\cite{c6,c7}. When the tip-sample separation(1 $\mu$m $< d < $ 10 $\mu$m) is well within the interaction range but still much larger than the modulation amplitude ( $\sim$ 10 nm), the tip-sample coupling is purely capacitive, providing a convenient and accurate way to align the two output channels.  We tune the reference phase to zero the lock-in reading in the MIM-R channel, whereas the MIM-C signal is automatically maximized. We have tested this alignment by setting the reference phase with a dielectric sample before approaching a lightly-doped Si wafer (10 $\Omega \cdot$cm or 10 S/m). The MIM-R channel only measures appreciable signals within a tip-sample spacing around 1 $\mu$m. The peak MIM signals in the approaching curve also contain important information about the local electrical properties\cite{c11}, which will be described in the next section.

\section{Results and Discussions}

\begin{figure}
	\centering
		\includegraphics[height=2.5in]{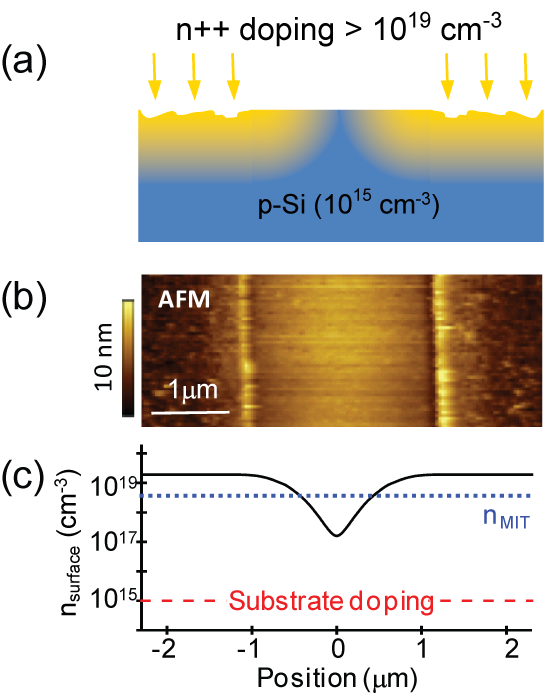}
		\caption{(color online) (a) Sketch of the cross section of the Si sample after ion-implantation and dopant diffusion. (b) AFM image of the sample surface, where the implanted region is slightly roughened due to the heavy dosage. (c) Simulated doping profile across the sample surface. The substrate background doping is also indicated (dash line). The dotted line indicates the metal-insulation transition density of doped Si. }
			\label{fig2}
\end{figure}

In this section, we demonstrate the performance of the cryogenic MIM setup using a selectively doped Si sample \cite{c21,c14}.  The schematic of the sample structure is illustrated in Fig. \ref{fig2}a. A dosage of 1 $\times$ 10$^{15}$ cm$^{-2}$ phosphorous was implanted into a stripe pattern on a lightly doped p-Si substrate (background doping 10$^{15}$ cm$^{-3}$ ), followed by a standard thermal annealing to activate the impurity atoms. The annealing process results in both lateral and vertical diffusion of the dopants. Fig. \ref{fig2}c shows the doping profile near the sample surface simulated by TSUPREM-4 software (Synopsys, Inc.).  The surface doping concentration decreases from $\sim$ 10$^{19}$ cm$^{-3}$ in the implanted region to $\sim$ 10$^{17}$ cm$^{-3}$ in the middle of the non-implanted region, which is still higher than the substrate doping level. Since the metal-insulator-transition (MIT) for phosphorous doped Si is approximately 3.7 $\times$ 10$^{18}$ cm$^{-3}$, a scan across the stripe will go through alternating metallic (low impedance) and insulating (high impedance) region at low temperatures. On the other hand, the ion implantation only causes minor damage on the sample surface, as shown in the atomic force microscope (AFM) image with RMS $\sim$ 1 nm (Fig. \ref{fig2}b).  In fact, the slight convolution of the surface roughness into the MIM images can be used to identify the implanted regions. The MIM contrast is almost purely electrical because of the small surface roughness, which also minimizes the tip wearing during the scan. This sample is therefore ideal for proof-of-principle impedance imaging experiments. Finally, the local conductivity and carrier concentration can be controlled by external parameters such as temperatures and electric fields. The well-documented properties of Si allow direct comparison between the experiment observation and theoretical expectations. 

\begin{figure}
	\centering
		\includegraphics[height=3.5in]{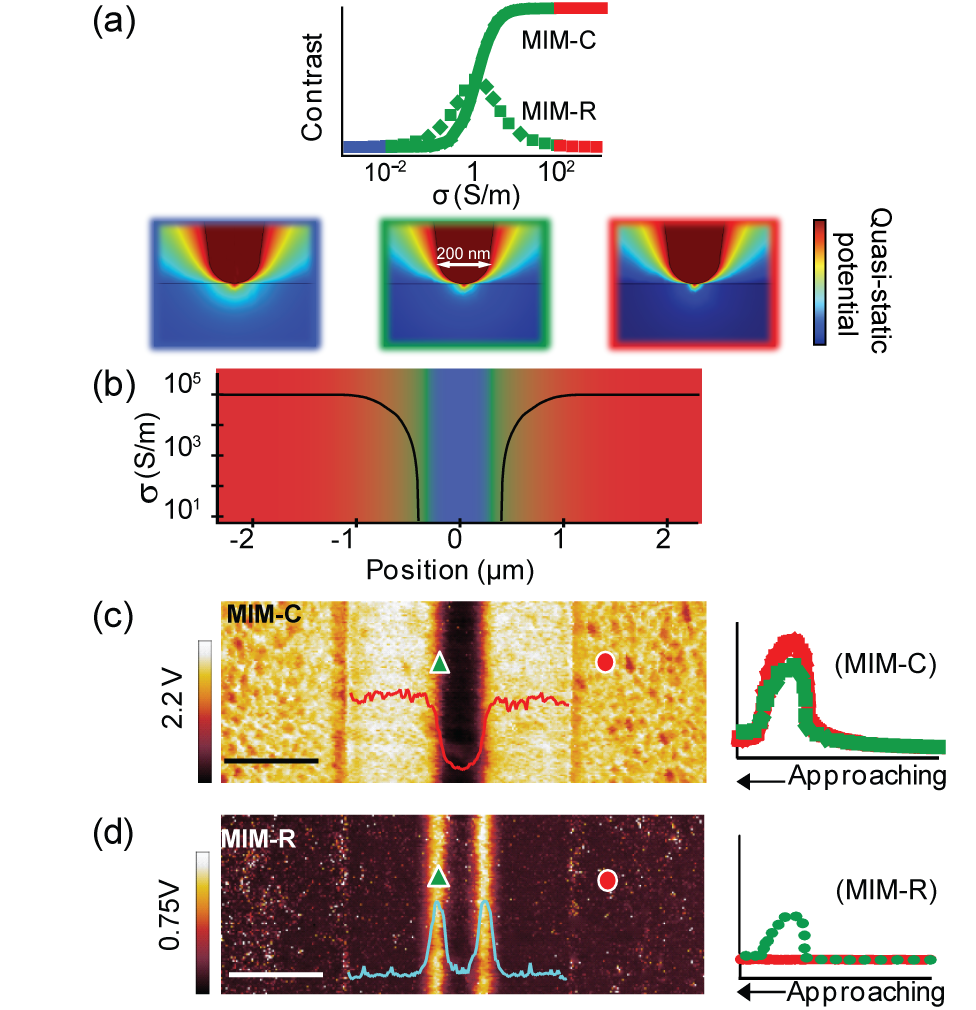}
		\caption{(color online) (a) The simulated MIM-C (solid) and MIM-R (dotted) contrast as a function of the sample conductivity. Quasi-static potential distributions around the tip electrode corresponding to the insulating (blue), conductive (red) and the crossover (green) regions are shown (see text for details). (b) Surface conductivity profile across the sample with the same color codes of the three regions. (c) MIM-C and (d) MIM-R images at 4 K. Typical line cuts are shown in each image. Approaching curves at two points (green triangle and red circle) are plotted in dotted and solid lines on the right side of (c) and (d).  All scale bars are 1 $\mu$m.  }
			\label{fig3}
\end{figure}

Before showing the MIM data, we briefly discuss the expected MIM responses across different regions of the sample surface. The modeling is performed using a finite element analysis (FEA) software\cite{c17} COMSOL 3.5 in the quasi-static and time harmonic ($f = 1$ GHz) mode \cite{c14}.  For this sample, a fully quantitative FEA simulation requires the complete knowledge of the 3D conductivity distribution due to diffusion of the dopants. In order to capture the essential physics without extensive computational effort, we consider here a much simplified approximation, in which only the local conductivity right underneath the tip is taken into account. The sample is assumed to be a semi-sphere, 0.5 $\mu$m in radius, with uniform conductivity in the insulating Si background \cite{c23}.  The tip (100 nm in radius) to ground admittance ($Y$, inverse impedance) as a function of $\sigma$ is calculated. The real and imaginary parts of $\Delta Y$ $=$ $Y_\sigma$ - $Y_0$, where $Y_0$ is the admittance of the background, is proportional to the contrast in MIM-R and MIM-C channels. Fig. \ref{fig3}a (top) shows the MIM-C and MIM-R curves for 10$^{-4}$ $< \sigma <$ 10$^4$ S/m. The curves are divided into three regimes\cite{c6,c7} and the potential distributions for $\sigma$ in these regimes are also shown (Fig. \ref{fig3}a, bottom). For $\sigma <$ 10$^{-2}$ S/m, the sample is in the insulating limit (blue) and both signals are small. The MIM-C contrast increases monotonically throughout the crossover region (green) and saturates in the conducting limit (red), $\sigma >$ 10$^2$ S/m. On the other hand, the MIM-R contrast is finite for 10$^{-2}$ $< \sigma <$ 10$^2$ S/m, reaching a maximum at $\sigma =$ 1 S/m and dropping back to zero in the conducting limit. We emphasize that the quasi-static potential distribution is insensitive to the height of the pyramidal part of the tip. It is well confined within several tip radii in the insulating limit and even better restricted underneath the tip in the conducting limit, confirming the assumption of local tip-sample interaction. Our simulation provides a semi-quantitative method for relating the MIM images to the local $\sigma$ at the sample surface.

Based on the surface doping profile in Fig. \ref{fig2}c, the expected surface conductivity profile at 4 K is plotted in Fig. \ref{fig3}b (left) \cite{c24}.  The sharp drop of $\sigma$, appearing 0.4 $\mu$m away from the center of the non-implanted region, corresponds to the well known metal-insulator transition (MIT) density in doped Si. Such an Anderson-Mott transition \cite{c25}, in which disorder effects and electron-electron interactions are equally important, was extensively studied using bulk-doped samples. Here, however, we can spatially resolve the transition on the sample surface with continuous density variation. In Fig. \ref{fig3}b, we overlay the same color labels of the simulation results (Fig. \ref{fig3}a) onto the surface conductivity profile. Good agreements between this semi-quantitative model and the actual MIM-C (Fig. \ref{fig3}c) and MIM-R (Fig. \ref{fig3}d) images are obtained. In Fig. \ref{fig3}c, the bright (conducting) areas extend from the implanted into the non-implanted regions and indeed stop sharply about 0.4 $\mu$m away from the center. At the same locations, two bright lines in the MIM-R image can be seen, indicating a transition from metallic ($\sigma >$ 10$^2$ S/m at low-T) to insulation ($\sigma <$ 10$^{-2}$ S/m at low-T) behaviors here. The width of the transition in both images is $\sim$ 0.15 $\mu$m, which is likely to be limited by the spatial resolution rather than the sharpness of the MIT itself. On the right of the MIM images in Fig. \ref{fig3}c and \ref{fig3}d, two sets of approaching curves recorded by the lock-in amplifiers are shown. As expected, the peak MIM-C value is bigger in the metallic region than that in the insulating region. On the other hand, a finite MIM-R peak is observed when approaching the point with density close to $n_{MIT}$, while the curve is flat on the degenerately doped area, confirming the metallic (lossless) behavior here. 

\begin{figure}
	\centering
		\includegraphics[height=2.2in]{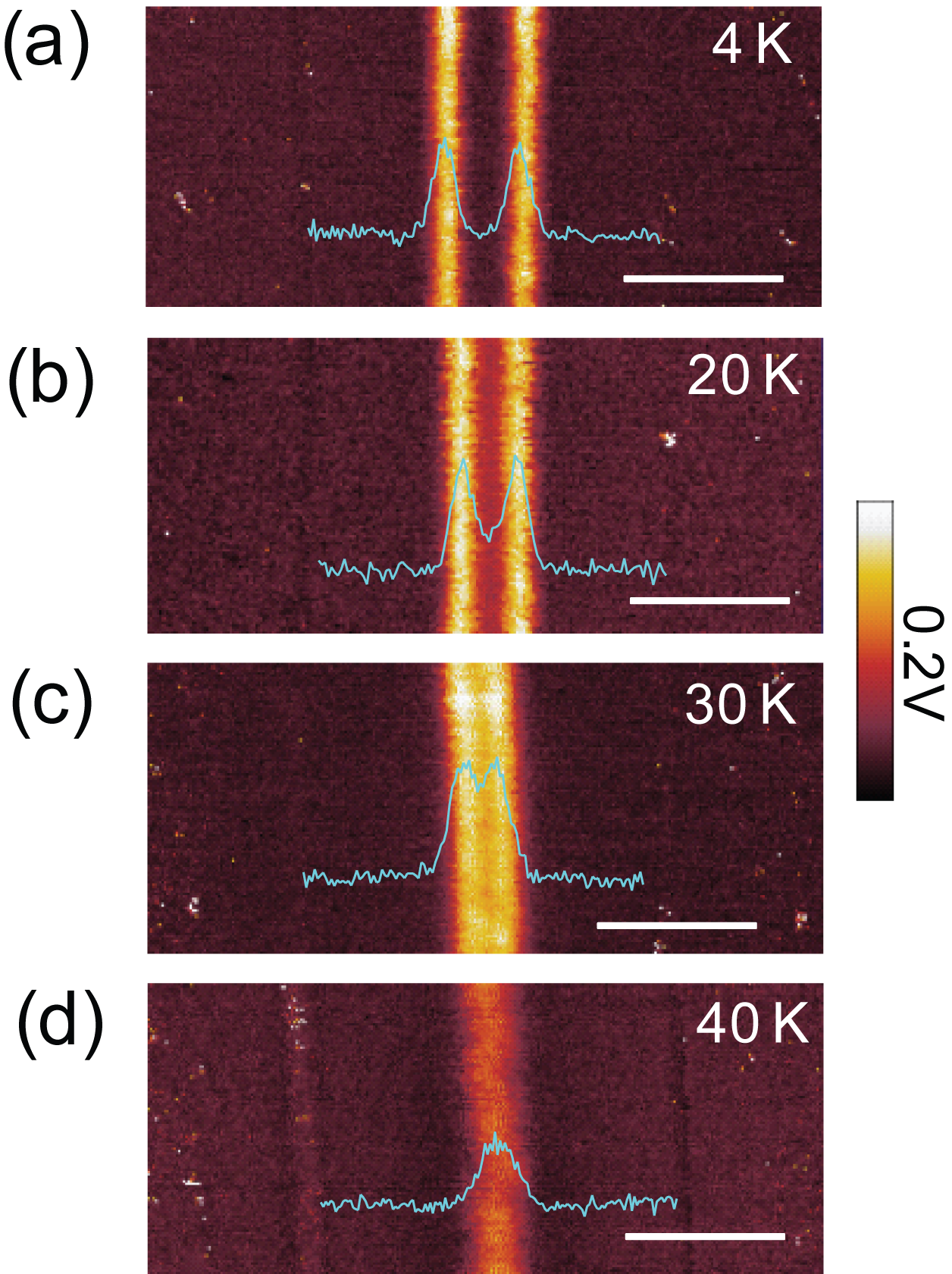}
		\caption{ (color online) (a) - (d) MIM-R images at different temperatures. Insets show typical line cuts for each image. All scale bars are 1 $\mu$m. }
			\label{fig4}
\end{figure}

Fig. \ref{fig4} shows the T-dependence of MIM images.  For increasing temperatures, the carriers are thermally excited in the non-degenerate region, resulting in enhanced local conductivity and reduced MIM contrast compared with the degenerate regions. This effect is best seen in the MIM-R images in Fig. \ref{fig4}a - \ref{fig4}d at 4, 20, 30, and 40 K, respectively. Typical line cuts of each image are also shown inside the images. At 20 K, the two bright lines move toward the center of the non-implanted region, where discernible MIM-R contrast with respect to the implanted regions starts to appear. At 30 K, while the two bright lines, corresponding to $\sigma \sim$ 1 S/m from the simulation, are still noticeable, the two peaks in the line cut almost merge with each other. At even higher temperatures, only one bright line is seen in the middle of the non-implanted region, the decrease of MIM-R contrast with increasing local conductivity at elevated T is consistent with the simulation results. The ability of imaging at variable temperatures, as demonstrated here, is highly desirable for the study of electronic phase transition in complex materials. 

\begin{figure}
	\centering
		\includegraphics[width=2.5in]{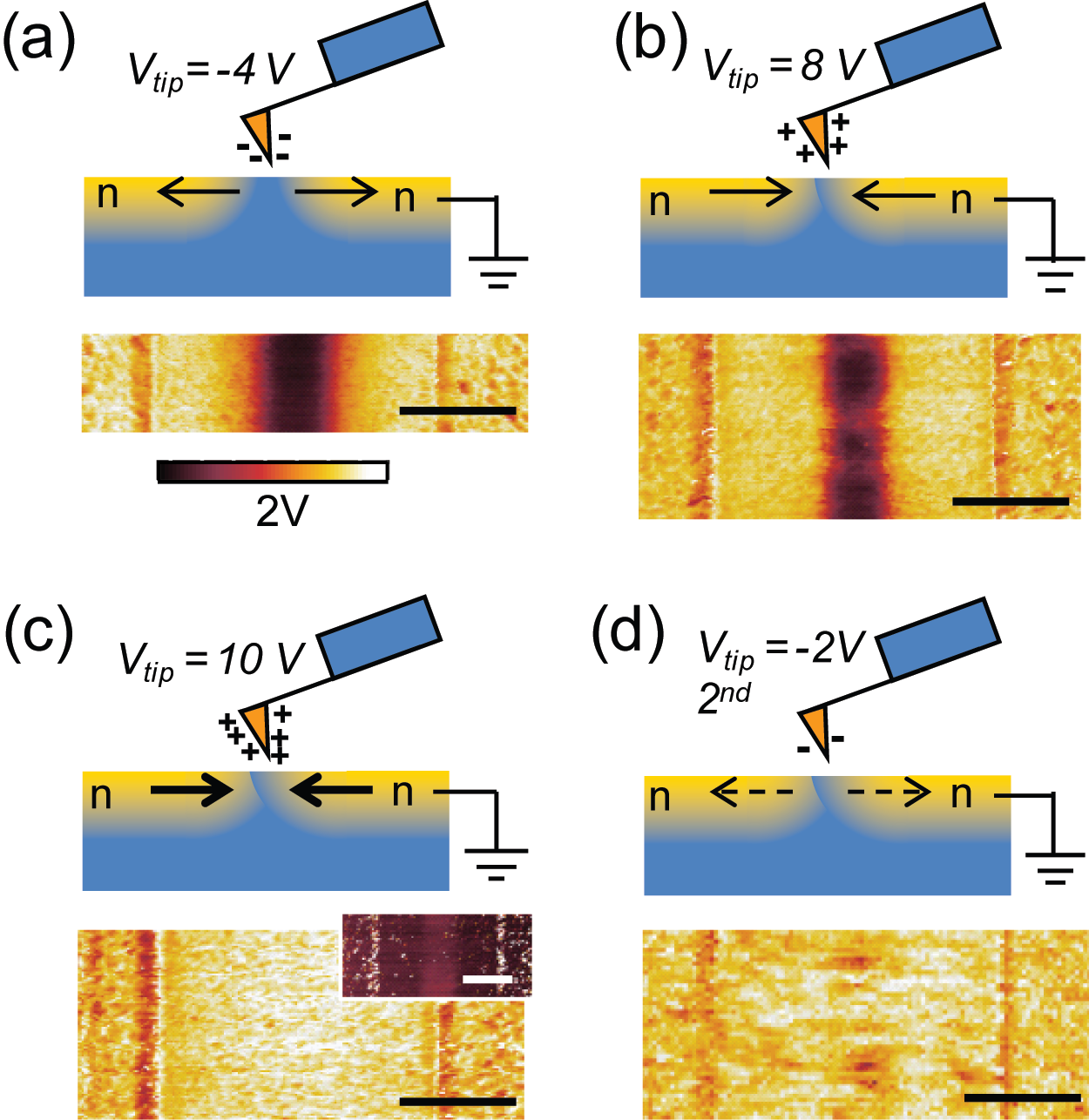}
		\caption{ (color online) (a) - (c): Effect of electric fields on the MIM images at $V_{tip}$ -4, 8 and 10 V, respectively. Cartoons of the carrier response to different DC biases are shown on top of the corresponding MIM-C images. The inset in (c) shows the MIM-R image at $V_{tip}$ = 10 V. (d) MIM-C image at -2 V after a previous scan with $V_{tip}$ = 10 V, showing the effect of carrier trapping at low temperatures. All scale bars are 1 $\mu$m.  }
			\label{fig5}
\end{figure}

External parameters such as electric\cite{c26, c27, c28}, and magnetic fields\cite{c29} are also available for impedance imaging in our system. For example, a DC voltage ($V_{tip}$) with respect to the sample can be applied to the tip through the bias tee (Fig. \ref{fig1}a).  The electric field induced by such voltage can repel or attract the n-type carriers. Unlike scanning capacitance microscopy, MIM measurement does not require an AC bias on the sample. Fig. \ref{fig5}a - \ref{fig5}c show schematics of the sample response and the corresponding MIM-C images taken at $V_{tip}$ = -4, 8, and 10 V, respectively.  For $V_{tip}$ = -4 V (Fig. \ref{fig5}a), the n-type carriers are repelled away from the tip, which widens the insulating region (dark).   For a positive bias, the n-type carriers are attracted into the non-implanted region to increase the local conductivity here.  At $V_{tip}$ = 8 V (Fig. \ref{fig5}b), the insulating region shrinks and the contrast versus the implanted region decreases. The effect is more prominent at $V_{tip}$ = 10 V, with vanishing contrast in both MIM-C and MIM-R channels (Fig. \ref{fig5}c and the inset). In fact, due to the low thermal energy at 4 K, the excess carriers are trapped at the same position even after the bias is removed. The MIM images remain almost the same after we gradually turned off the 10 V bias and applied a small negative $V_{tip}$ = -2 V, as seen in Fig. \ref{fig5}d. The equilibrium carries/conductivity distribution can be restored by a thermal cycle to room temperature (not shown). 

\section{Conclusion}

In summary, we describe the setup and the operation of a variable temperature microwave impedance microscope. The distance modulation mode plays an important role for the tip-sample approach, the sample plane fitting, and the reference phase alignment. We demonstrate the capability of MIM imaging by measuring the metal-insulator transition in a selectively doped Si sample. At 4 K, clear contrast between the degenerated and non-degenerated regions is observed in both channels. At elevated temperatures, the local conductivity is increased, pushing the transition towards the center of the non-implanted stripes. The effect of an electric field can be also studied by applying a bias on the MIM probe during imaging. Our results have demonstrated the ability to metal-insulator transitions and complex materials at cryogenic temperatures.

\begin{acknowledgments}
This research is funded by Center of Probing the Nanoscale (CPN), Stanford University, NSF grant DMR-0906027 and DOE Contract DE-FG03-01ER45929-A011 for low temperature cryostat. This publication is also based on work supported by Award No. KUS-F1-033-02, made by King Abdullah University of Science and Technology (KAUST) under the global research partnership (GRP) program. CPN is an NSF NSEC, NSF Grant No. PHY-0425897. 
\end{acknowledgments}



\begin{thebibliography}{99}
\bibitem{c23} In this case the dimensions along the stripes and the junction depth are both much larger than the tip size. Since the quasi-static potential usually drops quickly beyond a distance close to the tip diameter, the exact dimensions in the simulation have little effect on the results.
\bibitem{c29} The metal-insulator transition density also weakly depends on the external magnetic fields. However, due to the finite spatial resolution of the tip, we do not observe appreciable shift of the feature under 9 T and have not discussed the data here.
\end{thebibliography}
\end{document}